\documentclass[aps,twocolumn,showpacs,amsmath,amssymb,pra,superscriptaddress,floatfix,longbibliography]{revtex4-1}
\usepackage{braket}
\usepackage{amsmath}
\usepackage{amssymb}
\usepackage{mathtools}
\usepackage{graphicx}
\usepackage{dcolumn}
\usepackage{bm}
\usepackage{multirow}
\usepackage{leftidx}
\usepackage{color}
\usepackage{float}
\usepackage{xcolor}
\usepackage{soul}

\usepackage{amsfonts}
\usepackage{eufrak}
\usepackage[german,english]{babel}

\begin{document}
\title{Theoretical study of early time superradiance for atom clouds and arrays}
\author{F.~Robicheaux}
\email{robichf@purdue.edu}
\affiliation{Department of Physics and Astronomy, Purdue University, West Lafayette,
Indiana 47907, USA}
\affiliation{Purdue Quantum Science and Engineering Institute, Purdue
University, West Lafayette, Indiana 47907, USA}

\date{\today}

\begin{abstract}
We explore conditions for Dicke superradiance in a cloud of atoms by
examining the Taylor series expansion of the photon emission rate
at $t= 0$. By defining superradiance as an increasing photon emission
rate for $t\sim 0$, we have calculated the conditions for superradiance
for a variety of cases. We investigate superradiance as defined for
photon emission into all angles as well as directional superradiance
where the photon emission is only detected in a particular
direction. Although all of the examples are for two level
atoms that are fully inverted at $t=0$, we also give equations for
partially inverted two level atoms and for fully inverted multilevel
atoms. We give an algorithm for efficiently evaluating these equations for
atom arrays and determine superradiance conditions for large atom
number.
\end{abstract}

\maketitle

\section{Introduction}

Superradiance\cite{RHD1954,RH11971,GH11982,allen1987optical}
is a collective phenomenon where multiple
atoms radiate faster than that of $N$ individual atoms due to their
interaction through the quantized electromagnetic field. For atoms
within a small region, the peak photon emission rate can scale like
$N^2$ rather than $N$ from uncorrelated atoms. This leads to a burst
of radiation on a time scale much smaller than the radiative lifetime
of a single atom. There has been an extensive number of experimental studies of
superradiance in atoms\cite{SHM1973,GFP1976,GVH1977,JM11979,GGF1979,CLP1981,MGG1983,WYC2007,SBK2007,DBW2008,CPD2011,GHH2015,ZRJ2016,GAK2016,RKH2016,AKK2016,NT12016,GCB2017,SBF2017,CWM2018,LWH2019,MFO2020,FGR2021},
investigations of superradiance for a wide variety of other platforms\cite{MAV1997,SSW2007,RSS2010,CZW2016,MBS2016,BJV2017,RBB2018,GIF2019},
and many theoretical investigations of different cases.
\cite{FHM1972,FH11974,MF11976,SM11989,JQ11995,CKK2000,CHS2003,SFO2006,SC12008,AGK2008,SCS2010,BGB2010,LY12012,KP12017,KK12017,SR32017,CKB2018,KK12018,KRK2019,OMG2019,RTS2019,SC12019,SMG2020,DLY2020,CLL2020,FVB2021,LM12021,DM12021,HMR2021,MAG2021}

In this paper, we follow the philosophy of Ref.~\cite{MAG2021} to use
the early time behavior of the photon emission rate to determine whether or not
a group of
$N$ atoms are superradiant. They used the statistics of the first
two photons emitted to classify the system. For this they used the
$g^{(2)}(0)$ with the condition that superradiance occurs when the first
photon emission enhances the rate of the second photon emission,
i.e. $g^{(2)}(0)>0$. With this
insight, they do not need to solve for the time dependence of the master
equation to determine superradiance. They classified whether or not
a system is superradiant by using the variance of the eigenvalues of
the decay matrix, $\Gamma_{nm}$ in Eq.~\eqref{eq:gdef1} below, because the
$g^{(2)}(0)$ can be found in terms of these values. Also important,
this formulation allows for
interpretation of results from large systems

Instead of using $g^{(2)}(0)$, in this report,
we use the early time behavior of the
photon emission rate to determine whether or not a system is superradiant.
The idea is that the photon emission rate, $\gamma (t)$,
is an increasing function of
time at $t= 0$ when the system is superradiant. Using the criterion
$\dot{\gamma}(0)>0$, we obtain the superradiance condition in terms of
the trace of the square of the decay matrix. Because the decay matrix
is real and symmetric, this criterion is exactly that of Ref.~\cite{MAG2021}.
An advantage of the new definition is that it is computationally
more efficient so that larger systems can be investigated. It also leads to
different insight into the conditions for superradiance. Another advantage
is that this formulation allows for generalization to multilevel systems, systems
that are not fully inverted, directional superradiance, and efficient
evaluation for arrays.

In many of the examples below, we emphasize the difference between superradiance
as defined from the total emission rate and superradiance as defined as
the emission rate into a particular direction. As an example, Ref.~\cite{FGR2021}
experimentally demonstrated cases with clear superradiance in a particular
direction (as evidenced by increasing photon emission rate at early time)
with no enhancement in other directions and Ref.~\cite{HMR2021} theoretically
investigated directional superradiance for weakly illuminated atom
arrays. While the theory can be used for
randomly placed atoms, all examples below are for atom arrays. For two-
and three-dimensional arrays, we give examples of directional superradiance
with atom separations comparable to or larger than the wavelength of the
light for experimentally accessible atom numbers. We also take advantage
of the efficiency in evaluating $\dot{\gamma}(0)$ to demonstrate its
scaling with number of atoms and show that superradiance occurs
in two- and three-dimensional atom arrays for sufficient atom number.

In Sec.~\ref{sec:BT}, we give the basic theory when the atoms are approximated
as two level systems. In Sec.~\ref{sec:Early}, we derive the expressions
for the early time behavior of the photon emission rate for two level
atoms which can then be
used to define superradiance while Sec.~\ref{sec:Early2} is the early time
behavior for a specific type of multilevel atom. In Sec.~\ref{sec:Examp},
we explore several examples. In Sec.~\ref{sec:Summ} is a summary of these
results. The App.~\ref{sec:AppEquiv} gives a brief derivation of the
equivalence of our superradiance condition with that in Ref.~\cite{MAG2021}.

\section{Basic Theory: 2 states}\label{sec:BT}

In this section,
we are using an excitation scheme where the atomic structure is approximated
as a two level system. The atoms will be considered as fixed in space which
means we are ignoring the atom recoil.

\subsection{Master equation formalism}

All of the equations will use a
simplified notation to reduce their size.
For the $n$-th atom, the ground and excited states are $|g_n\rangle$ and
$|e_n\rangle$. The operators used below follow the definition
\begin{equation}
\hat{e}_n\equiv  |e_n\rangle\langle e_n|\qquad
\hat{\sigma}^-_n\equiv |g_n\rangle\langle e_n|\qquad
\hat{\sigma}^+_n\equiv|e_n\rangle\langle g_n|
\end{equation}

The equation for the $N$-atom density matrix\cite{RL12000} can be written in
the form
\begin{equation}\label{eq:DenMat}
\frac{d\hat{\rho}}{dt}=\sum_n\left(
{\cal L}_n(\hat{\rho })
+\sum_{m\neq n}\left[\frac{1}{i\hbar}[H_{nm},\hat{\rho}]
+{\cal L}_{nm}(\hat{\rho})
\right]\right)
\end{equation}
where the the one atom Hamiltonian that arises from an external
laser interacting with each atom is zero
for all of the examples below and has not been
included, ${\cal L}_n$ is from one atom decays
of Lindblad type, the $H_{nm}$ is the two atom Hamiltonian from the
dipole-dipole interactions, and ${\cal L}_{nm}$ are the two atom decays
from the dipole-dipole interactions. For the two level cases considered
here, these operators are
\begin{eqnarray}
{\cal L}_n(\hat{\rho })&=&
\frac{\Gamma}{2}(2\hat{\sigma}_n^-\hat{\rho}\hat{\sigma}_n^+-\hat{e}_n
\hat{\rho}-\hat{\rho}\hat{e}_n)\\
H_{nm}&=&\hbar\Omega_{nm}\hat{\sigma}^+_n\hat{\sigma}_{m}^-\\
{\cal L}_{nm}&=&\frac{\Gamma_{nm}}{2}(2\hat{\sigma}_n^-\hat{\rho}
\hat{\sigma}_{m}^+-\hat{\sigma}_{m}^+\hat{\sigma}_n^-
\hat{\rho}-\hat{\rho}\hat{\sigma}_{m}^+\hat{\sigma}_n^-)
\end{eqnarray}
where 
$\Gamma$ is the decay rate of a single atom.
The two atom parameters are defined for $m\neq n$ as
\begin{eqnarray}
\Gamma_{nm}&=&g(\boldsymbol{R}_{nm})+g^*(\boldsymbol{R}_{nm})=
2\Re [g(\boldsymbol{R}_{nm})]\label{eq:gdef1}\\
\Omega_{nm}&=&\frac{g(\boldsymbol{R}_{nm})-g^*(\boldsymbol{R}_{nm})}{2i}=
\Im [g(\boldsymbol{R}_{nm})]\label{eq:gdef2}\\
g(\boldsymbol{R}) &=&\frac{\Gamma}{2}\left[h_0^{(1)}(s)+
\frac{3\hat{R}\cdot\hat{d}^*\hat{R}\cdot\hat{d}-1}{2}
h_2^{(1)}(s)\right]\label{eq:gdef3}\\
g^\pm_{nm} &\equiv &\pm i\Omega_{nm} + \frac{1}{2}\Gamma_{nm}\label{eq:gdef4}
\end{eqnarray}
with $\boldsymbol{R}_{nm}=\boldsymbol{R}_n-\boldsymbol{R}_m$
with $\boldsymbol{R}_n$ the position of atom $n$,
$\hat{d}$ the dipole unit vector,
$s=kR$, $\hat{R}=\boldsymbol{R}/R$, and the $h_\ell^{(1)}(s)$ the outgoing spherical
Hankel function of angular momentum $\ell$:
$h_0^{(1)}(s)=e^{is}/[is]$ and $h_2^{(1)}(s) = (-3i/s^3 - 3/s^2 + i/s)e^{is}$. 
The $g(\boldsymbol{R})$ is proportional to the propagator that gives the
electric field at $\boldsymbol{R}$ given a dipole at the origin\cite{JDJ1999}.
For
a $\Delta M =0$ transition, $\hat{d}=\hat{z}$ and the coefficient
of the $h_2^{(1)}$ Bessel function is $P_2(\cos (\theta ))=
(3\cos^2(\theta )-1)/2$ where $\cos(\theta )=Z/R$. For
a $\Delta M =\pm 1$ transition, the coefficient
of the $h_2^{(1)}$ Bessel function is $-(1/2)P_2(\cos (\theta ))=
(1-3\cos^2(\theta ))/4$.
To simplify some formulas below, we will define the diagonal component
of $\Gamma_{mn}$ as
\begin{equation}
\Gamma_{nn}=2\Re [g(\boldsymbol{R}\to 0)]=\Gamma .\label{eq:gdef1p}
\end{equation}

\subsection{Photon emission rate}

The rate that photons are emitted into all angles at time $t$ is given
by
\begin{equation}\label{eq:GamTot}
\gamma (t)=\sum_n\left[\Gamma\langle\hat{e}_n\rangle (t)+\sum_{m\neq n}\Gamma_{mn}\langle\hat{\sigma}^+_m\hat{\sigma}^-_n\rangle (t)\right].
\end{equation}
The rate that photons are emitted into the $\hat{k}_f$ direction
is proportional to\cite{allen1987optical}
\begin{equation}\label{eq:GamDir}
\gamma(t,\boldsymbol{k}_f)=\Gamma \sum_n\left[\langle\hat{e}_n\rangle (t)+\sum_{m\neq n}
e^{i\varphi_{mn}}
\langle\hat{\sigma}^+_m\hat{\sigma}^-_n\rangle (t)\right]
\end{equation}
where $\varphi_{mn}={\boldsymbol k}_f\cdot(\boldsymbol R_m-\boldsymbol R_n)$
with
$\boldsymbol{k}_f=2\pi /\lambda_0 \, \boldsymbol{\hat{k}}_f$. The normalization
of $\gamma(t,\boldsymbol{k}_f)$ was chosen so that a fully inverted
system has $\gamma (0,\boldsymbol{k} )=N\Gamma$ in analogy with 
$\gamma (0)$. The definition Eq.~\eqref{eq:GamDir} only makes sense
if the orientation of the dipoles are not in the $\hat{k}$ direction
because the actual directional rate involves the direction of the
dipole and the $\hat{k}$.

An interesting question arises from these two definitions. As
noted by Ref.~\cite{MAG2021}, a natural definition of superradiance
is when the emission of the first photon enhances the rate that the second
photon is emitted. This implies the rate of photon emission, Eq.~\eqref{eq:GamTot}, is
an increasing function of time at $t=0$ since an increasing $\gamma (t)$
means the many
atoms radiate faster as time develops. This can only
occur when pair correlations
$\langle\hat{\sigma}^+_m\hat{\sigma}^-_n\rangle (t)\neq 0$
develop in the gas because the $\sum_n\langle\hat{e}_n\rangle$ is
a decreasing function of time for undriven
atoms. As we will see below and was experimentally observed in
Ref.~\cite{FGR2021}, there are cases where
$\dot{\gamma} (0,\boldsymbol{k})>0$ for some directions $\hat{k}$ even though
$\dot{\gamma} (0)<0$. However, for $\dot{\gamma} (0,\boldsymbol{k})>0$,
there must be nonzero (and substantial) pair correlations developing in the atom
cloud even when
$\dot{\gamma} (0)<0$. We will call this case ``directional superradiance" to distinguish
it from the case where $\dot{\gamma}(0)>0$.

Both types of
superradiance are interesting because the gas has become correlated.
This can be seen from the development of an initially fully inverted system which
has $\langle\hat{\sigma}^\pm_n\rangle (t)=0$. Therefore, superradiance demands
$\langle\hat{\sigma}^+_m\hat{\sigma}^-_n\rangle (t)-
\langle\hat{\sigma}^+_m\rangle (t)\langle\hat{\sigma}^-_n\rangle (t)\neq 0$
implying nonnegligible pair correlations.

\subsection{Uncorrelated initial state}

In the following, we will examine how correlations develop when starting
from a completely uncorrelated but not necessarily fully inverted initial state
\begin{equation}\label{eq:Psi0}
\ket{\psi_{\rm i}}=\Pi_{\otimes n}[\cos(\alpha /2)\ket{g_n}+e^{i\boldsymbol k_{\rm i}\cdot\boldsymbol R_n}\sin (\alpha /2)\ket{e_n}] 
\end{equation}
where $\sin^2(\alpha /2)$ is the probability for an atom to be excited
and $\hat{k}_i$ gives a phase change across the atom cloud. This form
for the initial state would result when a cloud was subject to an intense
but short laser pulse in the limit that the pulse duration gets very
short.
We will examine superradiance as a function of both $\alpha$ and
$\hat{k}_i$. Superradiance can occur when the cloud is not fully inverted.
The interplay between $\hat{k}_i$ and the shape of the cloud can lead to
interesting effects, e.g. if the cloud is elongated in the $\hat{k}_i$ direction.

\section{Evaluation of early time photon rates: 2 states}\label{sec:Early}

\subsection{First derivative}

We will use a somewhat different approach from Ref.~\cite{MAG2021}
to exactly evaluate the early time behavior of the photon emission rates, $\gamma$'s.
The idea is based on performing a Taylor series expansion
\begin{equation}
\gamma (t) = \gamma (0) + \dot{\gamma} (0) t + \frac{1}{2}\ddot{\gamma}(0)t^2 + ...
\end{equation}
where $\dot{\gamma }(0)$ means the first derivative of $\gamma$ evaluated
at $t=0$, etc. Superradiance occurs when $\dot{\gamma}(0)>0$.

The evaluation of $\dot{\gamma}(0)$ simply requires the derivatives
of $\langle\hat{e}_n\rangle$ and $\langle\hat{\sigma}^+_m\hat{\sigma}^-_n\rangle$
\begin{equation}
\frac{d \langle\hat{e}_n\rangle}{dt}=-\Gamma \langle\hat{e}_n\rangle
-\sum_{m\neq n}(g_{nm}\langle \hat{\sigma }^-_m\hat{\sigma}^+_n\rangle
+g^*_{nm}\langle \hat{\sigma}^+_m\hat{\sigma }^-_n\rangle)
\end{equation}
and
\begin{eqnarray}
\frac{d \langle\hat{\sigma}^+_m\hat{\sigma}^-_n\rangle}{dt}&=&
-\Gamma \langle\hat{\sigma}^+_m\hat{\sigma}^-_n\rangle +
2\Gamma_{mn}\langle\hat{e}_m\hat{e}_n\rangle -g_{mn}\langle \hat{e}_m\rangle -\nonumber\\
&\null&g^*_{mn}\langle \hat{e}_n\rangle+\sum_{l\neq m,n}[g_{nl}(2\langle\hat{\sigma}^-_l\hat{\sigma}^+_m\hat{e}_n\rangle
- \langle \hat{\sigma}^-_l\hat{\sigma}^+_m\rangle )\nonumber\\
&\null &\qquad +g^*_{ml}(2 \langle\hat{\sigma}^+_l\hat{e}_m\hat{\sigma}^-_n\rangle -
\langle \hat{\sigma}^+_l\hat{\sigma}^-_n\rangle )
]
\end{eqnarray}
from Ref.~\cite{RS12021}.

Using the initial wave function, Eq.~\eqref{eq:Psi0}, all of the expectation
values can be evaluated at $t=0$:
\begin{equation}
\langle\hat{e}_n\rangle=\frac{1-\cos\alpha}{2} \quad \langle\hat{\sigma}^-_n\rangle =
\langle\hat{\sigma}^+_n\rangle^* =\frac{\sin\alpha}{2} e^{i\boldsymbol{k}_i\cdot
\boldsymbol{R}_n}
\end{equation}
with all of the other expectation values being products of these,
$\langle \hat{A}\hat{B}\rangle = \langle \hat{A}\rangle\langle\hat{B}\rangle$,
since the initial wave function is a product state.

For clarity,
we first give the result for a fully inverted gas, $\alpha = \pi$,
for the total photon emission rate:
\begin{equation}\label{eq:Gdotfi}
\dot{\gamma}(0)=-N\Gamma^2+\sum_{n,m\neq n}\Gamma_{mn}\Gamma_{nm}
=-2N\Gamma^2 + {\rm Tr}[\underline{\Gamma}\;\underline{\Gamma} ]
\end{equation}
where $\underline{ \Gamma}$ means the matrix of $\Gamma_{mn}$ and
${\rm Tr}[...]$ means the trace. This expression arises because
$\langle\hat{e}_n\rangle =1$ and $\langle\hat{\sigma}^\pm_n\rangle =0$.
Because $\underline{\Gamma }$ is a real, symmetric matrix, our condition
$\dot{\gamma}(0)>0$ is identical to Eq.~(3) of Ref.~\cite{MAG2021} which gives
the superradiance condition in terms
of the variance of the eigenvalues of $\underline{\Gamma }$;
see App.~\ref{sec:AppEquiv} below. While the
form Eq.~(3) of Ref.~\cite{MAG2021} has advantages as discussed there,
Eq.~\eqref{eq:Gdotfi} has the advantage of being computationally faster
(number of operations scaling like $N^2$ instead of $N^3$) and providing insight into
scaling with large atom numbers (discussed below). Also, as discussed below,
the number of operations scales like $N^1$ for arrays.
The Dicke model\cite{RHD1954},
$\Gamma_{mn}=1$, in Eq.~\eqref{eq:Gdotfi}
gives $\dot{\gamma}(0)=N(N-2)\Gamma^2$ which
is the result from Dicke's original derivation.

The fully inverted
gas for directional emission has
\begin{eqnarray}\label{eq:Gdotfik}
\dot{\gamma}(0,\boldsymbol{k}_f)&=&-2N\Gamma^2+\Gamma \sum_{mn}\Gamma_{mn}\cos\varphi_{nm}\nonumber\\
&=&-2N\Gamma^2+\Gamma{\rm Tr}[\underline{\Gamma}\;\underline{\cos\varphi}]
\end{eqnarray}
where $\varphi_{mn}=\boldsymbol{k}_f\cdot (\boldsymbol{R}_m-\boldsymbol{R}_n)$.
Because this expression only has one
power of $\Gamma_{mn}$ (which decreases like $1/|\boldsymbol{R}_m-\boldsymbol{R}_n|$),
the condition $\dot{\gamma}(0,\boldsymbol{k}_f)>0$ can be satisfied more easily
than Eq.~\eqref{eq:Gdotfi} if the $\boldsymbol{k}_f$ is in the correct direction.

The equations for partially inverted samples are somewhat more complicated due to the
survival of terms with raising and lowering operators. The total
decay rate gives
\begin{eqnarray}
\dot{\gamma}(0)&=&-N\Gamma^2\frac{1-c}{2}+\sum_n\sum_{m\neq n}[\frac{c(c-1)}{2}\Gamma_{mn}\Gamma_{nm}-\frac{s^2}{2}\Gamma_{mn}\nonumber\\
&\times &
\{\Gamma \cos (\eta_{mn})-c \sum_{l\neq n,m}(g_{nl}e^{i\eta_{lm}}+g^*_{ml}e^{-i\eta_{ln}})\}
]
\end{eqnarray}
where $c\equiv\cos\alpha$, $s\equiv\sin\alpha$, and 
$\eta_{mn}={\bf k}_i\cdot ({\bf R}_m-{\bf R}_n)$.
The directional decay rate is
\begin{eqnarray}
\dot{\gamma}(0,{\bf k}_f)&=&-N\Gamma^2\frac{1-c}{2}+\Gamma\sum_n\sum_{m\neq n}[\frac{c(c-1)}{2}\Gamma_{mn}\cos\eta_{nm}\nonumber\\
&\null &
- \frac{s^2}{4}\{ \Gamma_{nm}\cos (\eta_{mn})+\Gamma
\cos (\varphi_{mn}-\eta_{mn})\nonumber\\
&\null&+ce^{i\varphi_{mn}}\sum_{l\neq n,m}e^{i\eta_{mn}}(g_{nl}e^{i\eta_{lm}}+g^*_{ml}e^{-i\eta_{ln}})\} 
]
\end{eqnarray}
An important point to note for the partial inversion is that the number
of operations scales like $N^3$ so these are more difficult calculations.

\subsection{Second derivative: fully inverted}

The second derivative of the photon emission rates are relatively
straightforward to evaluate when the atoms are fully inverted using\cite{RS12021}
\begin{equation}
\frac{d^2\langle\hat{e_n}\rangle }{dt^2}(0)=\Gamma^2-\sum_{m\neq n}\Gamma_{nm}\Gamma_{mn}
\end{equation}
and
\begin{equation}
\frac{d^2\langle\hat{\sigma}^+_m\hat{\sigma}^-_n\rangle}{dt^2}(0)
=-4\Gamma\Gamma_{mn}+\sum_{l\neq m,n}(g_{nl}\Gamma_{lm}+g^*_{ml}\Gamma_{ln}).
\end{equation}
Using these expressions gives
\begin{eqnarray}
\ddot{\gamma}(0)&=&N\Gamma^3-5\Gamma\sum_{nm}(1-\delta_{nm})\Gamma_{nm}
\Gamma_{mn}\nonumber \\
&\null&+\sum_{nml}(1-\delta_{nl})(1-\delta_{ml})(1-\delta_{nm})\Gamma_{nm}
\Gamma_{ml}\Gamma_{ln}\nonumber \\
&=&8N\Gamma^3-8\Gamma {\rm Tr}[\underline{\Gamma}\; \underline{\Gamma}]
+{\rm Tr}[\underline{\Gamma}\; \underline{\Gamma}\; \underline{\Gamma}]
\end{eqnarray}
for the total decay rate.  The Dicke model\cite{RHD1954},
gives $\ddot{\gamma}(0)=N(N^2-8N+8)\Gamma^3$ which
is the result from Dicke's original derivation.
For the directional decay rate,
\begin{eqnarray}
\ddot{\gamma}(0,{\bf k}_f)&=&8N\Gamma^3-2\Gamma {\rm Tr}[\underline{\Gamma}\;
\underline{\Gamma}]-6\Gamma^2{\rm Tr}[\underline{\Gamma}\;
\underline{\cos\eta}]\nonumber\\
&\null&+\Gamma {\rm Tr}[\underline{\Gamma}\;\underline{\Gamma}\;
\underline{\cos\eta}]+\Gamma {\rm Tr}[\underline{\sin\eta}\; [\underline{\Gamma},\underline{\Omega}]\;
]
\end{eqnarray}

In principle, this logic can be continued to higher order derivatives. With
higher derivatives, it might be possible to determine the peak
fluorescence and the time at which it occurs. The higher derivatives will
require larger powers of the $\underline{\Gamma}$ which would be most efficiently
evaluated using the diagonalization described in Ref.~\cite{MAG2021}.

\section{Evaluation of early time photon rates: many final states}\label{sec:Early2}

This section gives the $\dot{\gamma}(0)$ when the excited state can decay
to several final states. To simplify the notation and derivation, we will
do the calculation without spin-orbit and hyperfine effects and further
assume the initial state has $\ell =0$. Extending beyond these restrictions
does not seem to be qualitatively different. We will denote the
principle quantum number of the excited state as
$\alpha_i$ and it can decay to many final states
with principle quantum number $\alpha_f$ with $\ell_f=1$. Instead of using $m_f$,
we will use Cartesian orbitals $i=x,y,z$.

The operators will be extended as
\begin{equation}
\hat{\sigma}^{\alpha_fi-}_n\equiv |(\alpha_fi)_n\rangle\langle e_n|
\qquad \hat{\sigma}^{\alpha_fi+}_n\equiv |e_n\rangle\langle (\alpha_fi)_n|
\end{equation}
with the $\hat{e}_n$ operator unchanged. Note the condition
\begin{equation}
\hat{\sigma}^{\alpha_f'i'+}_n\hat{\sigma}^{\alpha_fi-}_n=\hat{e}_n\delta_{ii'}
\delta_{\alpha_f\alpha_f'}
\end{equation}
The Lindblad term, with $n=m$ allowed, is
\begin{eqnarray}
{\cal L}(\hat{\rho})&=&\frac{1}{3}\sum_{nm\alpha_fii'}\frac{\Gamma_{nm}^{\alpha_fii'}}{2}
(2\hat{\sigma}^{\alpha_fi-}_n\hat{\rho}\hat{\sigma}^{\alpha_fi'+}_m\nonumber\\
&\null &-\hat{\sigma}^{\alpha_fi'+}_m\hat{\sigma}^{\alpha_fi-}_n\hat{\rho}
-\hat{\rho}\hat{\sigma}^{\alpha_fi'+}_m\hat{\sigma}^{\alpha_fi-}_n)
\end{eqnarray}
where
\begin{equation}
\Gamma_{nm}^{\alpha_fii'} =\Gamma_{\alpha_f}\left[j_0(k_{\alpha_f}R)+
\frac{3\hat{R}_i\hat{R}_{i'}-1}{2}
j_2(k_{\alpha_f}R)\right]
\end{equation}
with $\Gamma_{\alpha_f}$ the total decay rate into state $\alpha_f$,
$k_{\alpha_f}$ the wave number of the photon that transitions from the
initial state to the $p$-state $\alpha_f$, $j_\ell(z)$ the usual spherical
Bessel functions,
$R=|{\bf R}_n-{\bf R}_m|$, and $\hat{R}=({\bf R}_n-{\bf R}_m)/R$.

Repeating the derivation of the previous section the slope of the photon
emission rate can be found.
The rate that photons of wave number magnitude $k_{\alpha_f}$ are emitted for a fully
inverted system is
\begin{equation}
\gamma_{\alpha_f}(0)=N\Gamma_{\alpha_f}.
\end{equation}
The slope of the photon emission for a fully inverted system is
\begin{equation}
\dot{\gamma}_{\alpha_f}(0)=-N\Gamma_{\alpha_f}\Gamma +
\sum_n\sum_{m\neq n}\frac{1}{9}\sum_{ii'}\left( \Gamma_{nm}^{\alpha_fii'}\right)^2
\end{equation}
where the total decay rate
$\Gamma =\sum_{\alpha_f}\Gamma_{ \alpha_f}$. Note that the initial
slope needs more atoms to have $\dot{\gamma}_{\alpha_f}(0)>0$ because
the negative term is relatively larger: the second term is proportional
to $\Gamma_{\alpha_f}^2$ while the first term is proportional to
$\Gamma_{\alpha_f}\Gamma$. The second term is proportional to
$N^2$ so adding more atoms in a compact region will lead to superradiance
even when $\Gamma \gg \Gamma_{\alpha_f}$, e.g. Rydberg states.

\section{Examples}\label{sec:Examp}

In this section, we discuss the results of calculations for several examples in
one-, two-, and three-dimensions.

\subsection{One-dimensional array}

In this section, we describe results for examples where the atoms are
in one or two lines with equal spacing between the atoms.
Our results for
the total decay rate in
a one-dimensional atom array match those of Ref.~\cite{MAG2021} and will not be
discussed in detail here. We will mainly focus on the directional photon
emission. We will restrict the dipole moment to be in the $z$-direction
and the atoms to be on one or two lines parallel to the $y$-axis.

\begin{figure}
\resizebox{86mm}{!}{\includegraphics{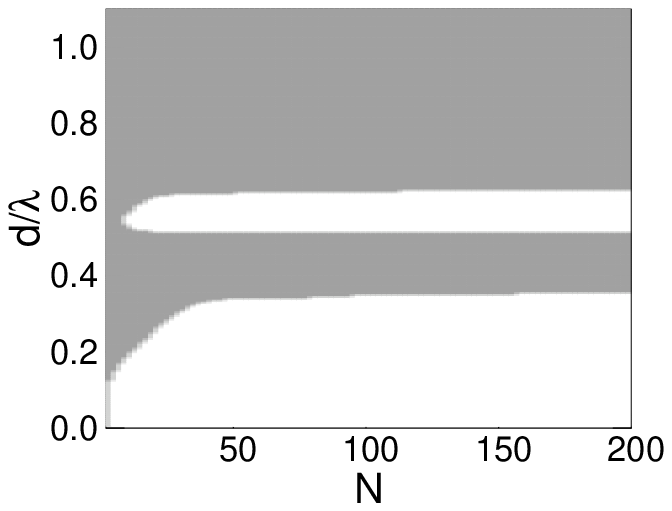}\includegraphics{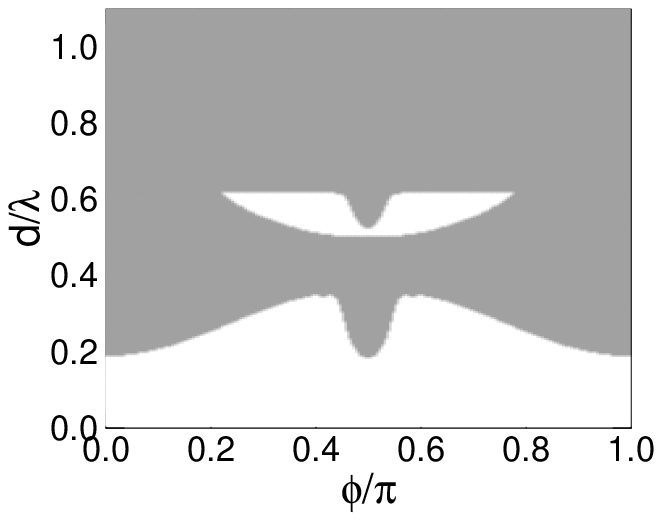}}
\caption{\label{fig:1D_1line}
Both plots are for an atom array in a line on the $y$-axis with the
atoms separated by $d$. The polarization is in the $z$-direction.
White shows the region where $\dot{\gamma}(0,{\bf k}_f)>0$ and gray
is where $<0$ with ${\bf k}_f=k(\hat{x}\cos\phi + \hat{y}\sin\phi )$.
The left plot has $\phi = 0.4\pi$. The right plot is for 100 atoms.
For the directional decay,
there is more than 1 region of superradiance unlike the case for
$\dot{\gamma}(0)$.
For the left
plot, the upper region only exists for
$N\geq 9$.
}
\end{figure}

As with Ref.~\cite{MAG2021}, there are regions where the slope of the
photon emission rate is larger than 0, indicating superradiance in
different directions. After fixing the polarization direction and the
line of atoms, there are three parameters of interest: the number
of atoms $N$, the separation of atoms $d$, and the angle of photon
emission. In Fig.~\ref{fig:1D_1line}, we show the region of superradiance
as defined by $\dot{\gamma}(0,{\bf k}_f)>0$ for
${\bf k}_f=k(\hat{x}\cos\phi + \hat{y}\sin\phi )$. The superradiant region
is white. In this case, there is a single line of atoms.
One plot shows the superradiant region as a function of $N,d$ for $\phi = 0.4\pi$
and the other shows this region versus $\phi ,d$ for $N=100$.
The plot versus $\phi ,d$ repeats for $\phi\to\phi +\pi$ and is symmetric
about $\phi = \pi /2$ due to the symmetry for a line of atoms in the
$y$-direction.

As with the results for the total emission rate in Ref.~\cite{MAG2021},
there is a region of relatively rapid change with $N$ for $N$ less than
about 20 followed by slower change with $N$ which apparently converges
to particular values for large $N$. The region of slow increase
is discussed in Sec.~\ref{sec:NDeff}. Unlike the total emission rate,
the directional emission has two regions of superradiance for larger
$N$ depending on the angle of emission. For $\phi$ between $\sim \pi /4$
and $\sim 3\pi /4$, there is a region of larger separation, $d/\lambda$
roughly between 0.5 and 0.6, where the photoemission rate increases with
time at early times. This is due to constructive interference in these
directions and is not present for the total emission rate which is only
superradiant for $d$ less than
$\simeq\lambda /4$ for $N=100$. This region of directional
superradiance for $d\sim \lambda /2$ is only present for $N\geq 9$ which
is not a large number of atoms. It seems possible to experimentally observe this
directional superradiance at larger $d$.

\begin{figure}
\resizebox{86mm}{!}{\includegraphics{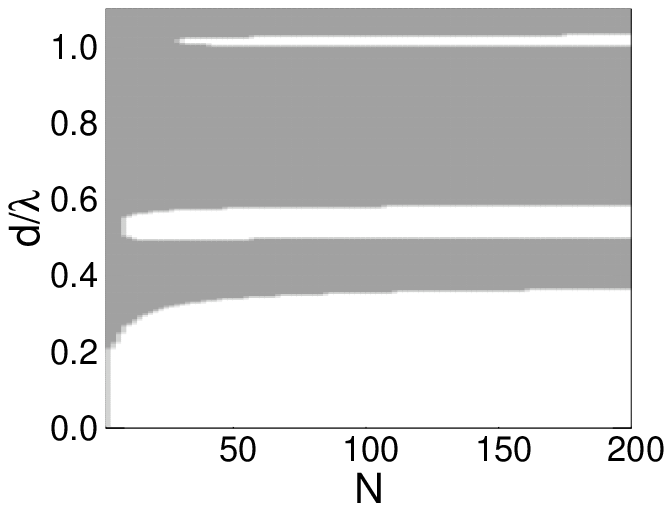}\includegraphics{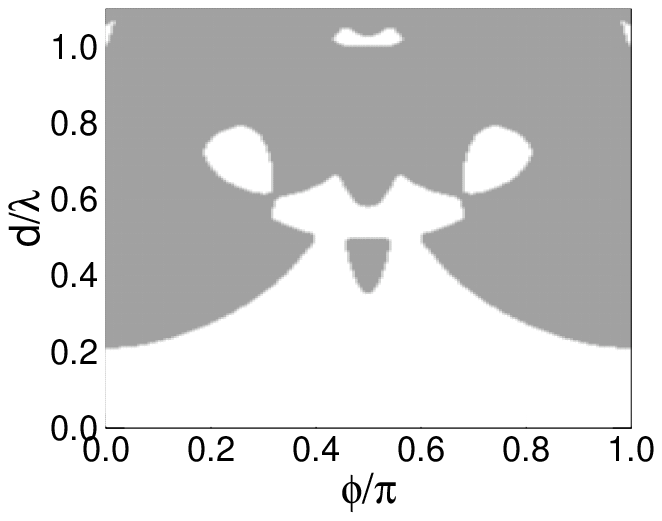}}
\caption{\label{fig:1D_2line}
Same situation as Fig.~\ref{fig:1D_1line}, except there are two lines
of atoms, each with $N/2$ atoms. The lines of atoms are displaced from
each other by
$d$ in the $x$-direction. For the left plot, $\phi = 0.5\pi$.
The middle region of superradiance only exists for $N\geq 9$ and
the upper region only exists for $N\geq 30$. Note the upper region is
superradiant for $d$ larger than $\lambda$.
}
\end{figure}

A more complex situation occurs if there are two, parallel lines of atoms.
Figure~\ref{fig:1D_2line}
shows results when the second line of atoms is displaced in the
$x$-direction by $d$. Each line contains $N/2$ atoms. For this case,
one plot shows the superradiant region as a function of $N,d$ for $\phi = \pi /2$
and the other shows this region versus $\phi ,d$ for $N=100$.
There are two interesting differences from the example with one line of
atoms. The first is that there are more regions of directional superradiance.
For $\phi =\pi /2$, there are three regions with the middle region
starting at $N=9$ and the upper region starting at $N=30$. For larger
$N$, there is a rich structure of superradiance on the $\phi ,d$ plane
due to the intereference between the different lines of atoms. The
second is that there is directional superradiance for $d>\lambda $ for
$\phi\simeq 0,\pi /2,\pi$. This region of superradiance for $d>\lambda$
requires more atoms, but $N$ is not so large that it is out of reach
for experimental investigation.

\subsubsection{Non-ideal cases}

For the cases in Figs.~\ref{fig:1D_1line} and \ref{fig:1D_2line}, we
calculated the effect of not fully inverting the atoms
for $N=100$ for $d<1.1\lambda$. For this case,
we assumed the laser propagation is in the $z$-direction giving
$\eta_{mn}=0$. As $\alpha$
decreases from $\pi$, the region of superradiance shrinks. When the
excited population decreases too much, the superradiance regions for
larger $d$, i.e. $d\sim\lambda /2$ for Fig.~\ref{fig:1D_1line}
and $d\sim \lambda /2$ and $\sim \lambda$ for Fig.~\ref{fig:1D_2line},
disappear. For the single
line case, the upper region disappears when the initial excitation population
decreases below 75\%. For the double line case, the region near
$d\sim\lambda $ disappears for less than approximately 80\% excited while the
region near $d\sim\lambda /2$ survives down to approximately 55\% excited.

Another possible non-ideal case has the atoms fully inverted but some of
the atoms are randomly removed which was
treated in Ref.~\cite{MAG2021}. We simulated this by randomly removing
each atom with a probability, $P$. For each $P$,
we repeated the simulations 100 times and
checked the superradiance condition. As with the non-fully inverted case,
the region of superradiance shrinks with increasing probability for
atom removal and at some point the superradiance regions for larger
$d$ disappear. For the single line case, the upper region disappears on average
when the number of atoms is less than 80\% while for the double line case
the region near $d\sim\lambda $ disappears for less than 70\% atoms while the
region near $d\sim\lambda /2$ survives down to approximately 40\% atoms.

Similar behavior is seen for higher dimensional arrays.

\subsection{Two-dimensional array}

In this section, we describe results for an example where the atoms are
in a two-dimensional array with equal spacing between the atoms.
Our results for the total decay rate in
a two-dimensional atom array match those of Ref.~\cite{MAG2021}; we discuss
these results below with those from a cubic array.
We restrict the dipole moment to be in the $z$-direction
and the atoms to be on a square array in the $xy$-plane of
size $N_1\times N_1$.

\begin{figure}
\resizebox{86mm}{!}{\includegraphics{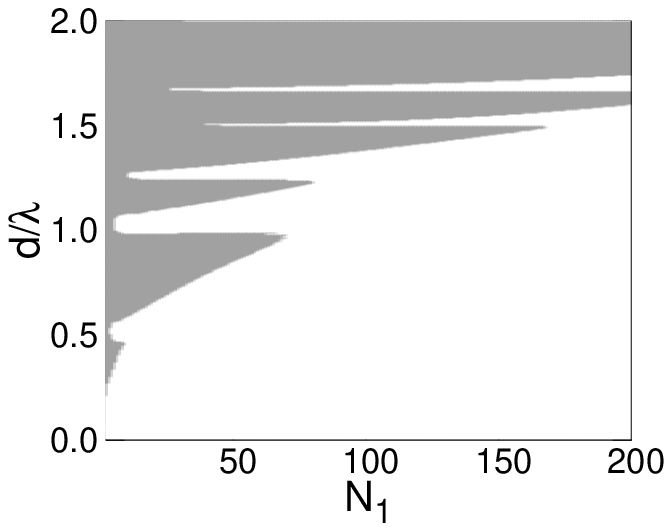}\includegraphics{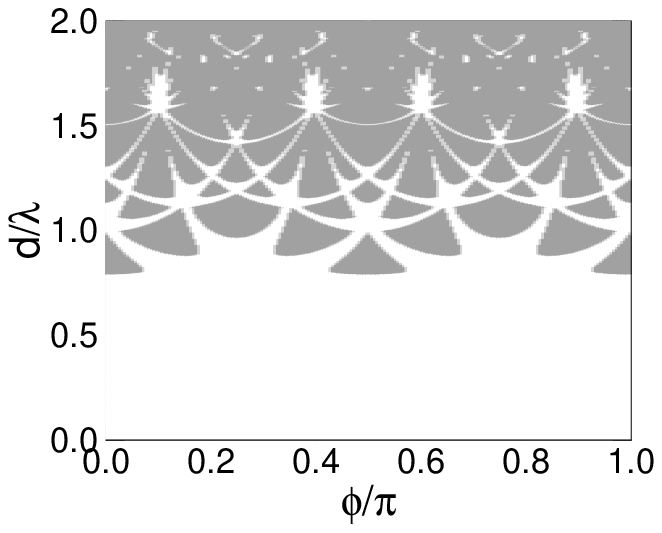}}
\caption{\label{fig:2D_1plane}
Same situation as Fig.~\ref{fig:1D_1line}, except there is a square array
of atoms in the $xy$-plane of size $N_1\times N_1$.
The dipole moment is perpendicular to the plane and
${\bf k}_f=k(\hat{x}\cos\phi + \hat{y}\sin\phi )$. For the left plot, $\phi = 0$
and the right plot has $N_1=40$.
}
\end{figure}

In Fig.~\ref{fig:2D_1plane}, we show the superradiant region versus
$N_1,d$ for $\phi =0$ and versus $\phi ,d$ for $N_1=40$ corresponding
to $N=1600$ atoms.
As with the results for the total emission rate in Ref.~\cite{MAG2021},
there is a region of relatively rapid change with $N_1$ for $N_1$ less than
about 20 followed by slower change with $N_1$. The change was even slower
in the plots of Ref.~\cite{MAG2021} because the plots were versus the
total number of atoms $N=N_1^2$ which greatly stretches the
abscissa. Unlike the one-dimensional case, it is
not clear whether the superradiant regions converge with increasing
$N_1$. It appears that the separation of atoms leading to
superradiance increases as the number atoms increases.
This is discussed in Sec.~\ref{sec:NDeff} where it is shown
the results do not converge with increasing $N$.

The directional superradiance for a plane is much richer than that for
one or two lines of atoms. In the plot versus $\phi ,d$ at $N_1=40$,
the results repeat for $\phi \to \phi +\pi /2$ and are symmetric
about $\phi = \pi /4$ and $3\pi /4$ because of the symmetry
for a square array. In addition, there are many more regions of
superradiance due to the different possible directions for constructive
interference. Some of these regions start at relatively small $N_1$. For
example, the superradiant region for $d\sim \lambda$ for $\phi =0$ starts
at $N_1=6$ corresponding to 36 atoms. The superradiant region for
$d\sim 5\lambda /4$ starts for $N_1=11$ corresponding to 121 atoms.
Both of these cases are within current experimental capabilities.\cite{RWR2020}
Perhaps even more interesting are the regions where different
constructive interference conditions overlap, for example, the regions
for $d\sim 1.6\lambda$ for $\phi \sim \pi /8, 3\pi /8, 5\pi /8, 7\pi /8$.
Also of interest are the superradiant regions where $d\sim 2\lambda$;
however, these regions are small and may not be robust to lattice
imperfections.

\subsection{Three-dimensional array}

In this section, we describe results for an example where the atoms are
in a three-dimensional array with equal spacing between the atoms.
We restrict the dipole moment to be in the $z$-direction
and the atoms to be on a cubic array of size $N_1\times N_1\times N_1$.

\begin{figure}
\resizebox{86mm}{!}{\includegraphics{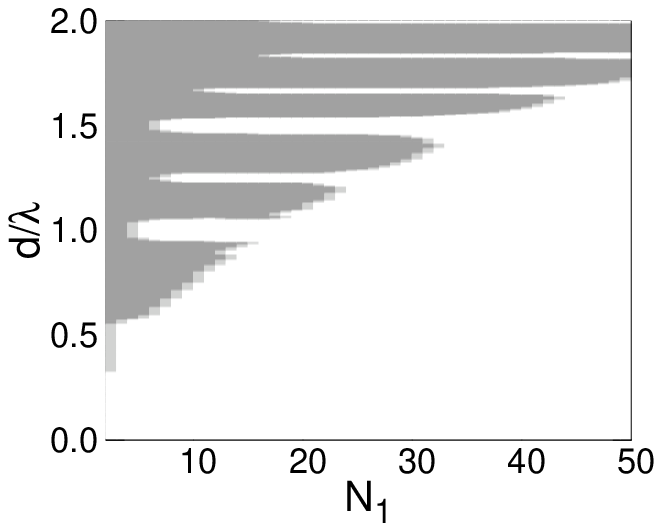}\includegraphics{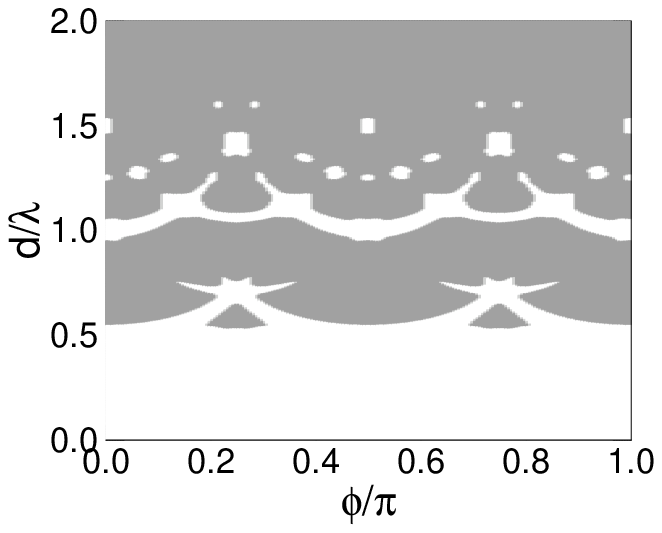}}
\caption{\label{fig:3D_1cube}
Same situation as Fig.~\ref{fig:1D_1line}, except there is a cubic array
of atoms of size $N_1\times N_1\times N_1$.
The dipole moment is in the $z$-direction and
${\bf k}_f=k(\hat{x}\cos\phi + \hat{y}\sin\phi )$. For the left plot, $\phi = 0$
and the right plot has $N_1=10$.
}
\end{figure}

We will first examine the case for directional superradiance. In Fig.~\ref{fig:3D_1cube},
we show the superradiant region versus
$N_1,d$ for $\phi =0$ and versus $\phi ,d$ for $N_1=10$ corresponding
to $N=1000$ atoms. As with the planar array, there is a richness to the
regions that arise due to directions giving constructive interference.
As with the planar array, in the plot versus $\phi ,d$ at $N_1=10$,
the results repeat for $\phi \to \phi +\pi /2$ and are symmetric
about $\phi = \pi /4$ and $3\pi /4$ because of the symmetry
for a cubic array. There are also several regions that are superradiant
for $d$ greater than $\sim \lambda$ for relatively small number of atoms.
Another interesting similarity to the two-dimensional case is the
increasing separation of atoms leading to
superradiance as the number atoms increases.
This is discussed in Sec.~\ref{sec:NDeff} where it is shown
the results do not converge with $N$.

\begin{figure}
\resizebox{86mm}{!}{\includegraphics{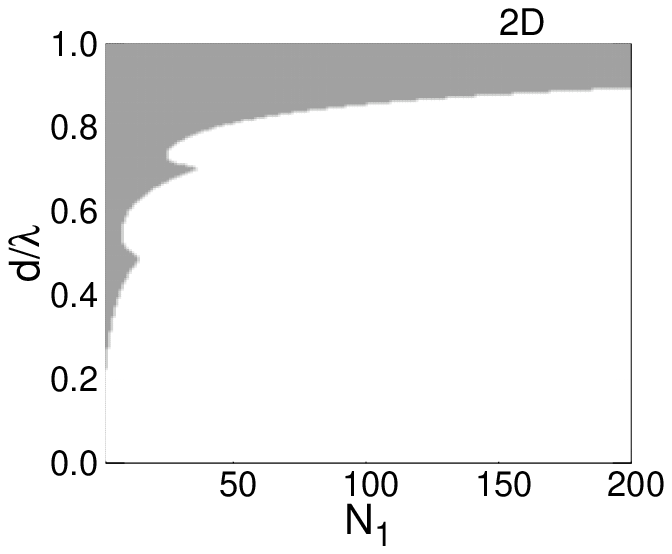}\includegraphics{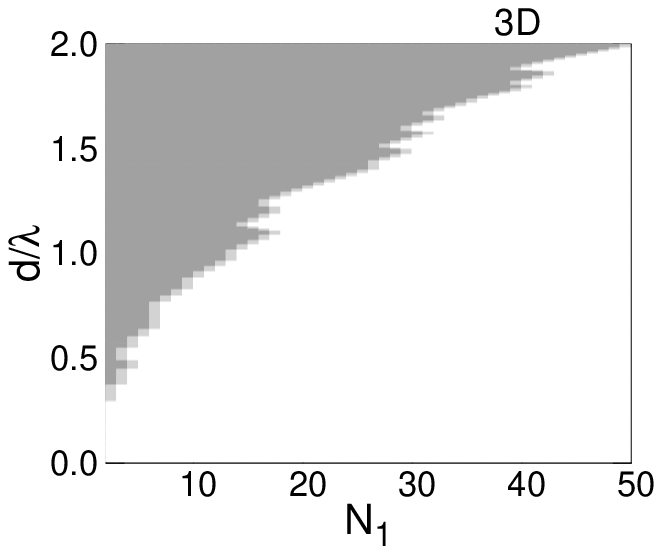}}
\caption{\label{fig:totND}
For the same case as Figs.~\ref{fig:2D_1plane} (for 2D) and \ref{fig:3D_1cube}
(for 3D),
plots of the superradiant region defined by $\dot{\gamma}(0)>0$.
}
\end{figure}

We also show the superradiant region for total photon emission,
$\dot{\gamma}(0)$, versus $N_1,d$ for the square and
cubic arrays in Fig.~\ref{fig:totND}. For the square array, the region
for $N_1\leq 40$ matches that shown in Ref.~\cite{MAG2021} Fig.~4
for the polarization perpendicular to the plane. Interestingly, the
maximum $d$ for superradiance appears to be an increasing function
of $N_1$ up to the largest values shown in Fig.~\ref{fig:totND}.
From Eq.~\eqref{eq:2Dasym},
the maximum $d/\lambda$ for superradiance is proportional
to $\sqrt{\ln{N_1}}$ which does continue to increase with $N_1$, albeit slowly.
For the cubic array, there are more regions of large separation superradiance
and they appear at smaller $N_1$. This is not surprising because there are
many more atoms close to each other which leads to faster radiation.
More interesting, from Eq.~\eqref{eq:3Dasym},
the maximum $d/\lambda$ for superradiance is proportional to $\sqrt{N_1}$.
Unlike the one-dimensional case, the two- and three-dimensional arrays do
not converge to superradiance properties as $N_1$ increases.

\subsection{Efficient summation and asymptotic trends}\label{sec:NDeff}

The summation in Eqs.~\eqref{eq:Gdotfi} and \eqref{eq:Gdotfik} require
$O(N^2)$ operations for a general positioning of atoms. While this is more
efficient than $O(N^3)$ operations, there are more efficient algorithms
in some cases. For example, for
an array, most of the terms are repeated. This fact can be used to reduce
the number of operations to $O(N)$.

For a single line,
there are $N$ terms where $n-m=0$; there are $N-1$ terms where $n-m=1$ or
$n-m=-1$ etc. This allows the calculation in terms of the difference
in positions and a weight for a given difference. As another example,
for a three-dimensional array with $N_1$ points in each direction
with lattice vectors ${\bf a}_1,{\bf a}_2,{\bf a}_3$ the atom positions can be
written as ${\bf R}_{n}=\nu_{1,n}{\bf a}_1 + \nu_{2,n}{\bf a}_2 + \nu_{3,n}{\bf a}_3$
where the $1\leq \nu_1,\nu_2,\nu_3\leq N_1$ are integers. There are
$\prod_i (N_1-|\nu_i|)$ terms that have
$\nu_{1,n}-\nu_{1,m}=\nu_1$, $\nu_{2,n}-\nu_{2,m}=\nu_2$,
and  $\nu_{3,n}-\nu_{3,m}=\nu_3$. Defining the scaled initial slope of
the photoemission rate as $\dot{\gamma}(0)/(N\Gamma^2)$,
this allows the reduction to
\begin{eqnarray}
\frac{\dot{\gamma}(0)}{N\Gamma^2}&=&-2+\sum_{\nu_1\nu_2\nu_3}W_{\nu_1\nu_2\nu_3}
\frac{\Gamma_{\nu_1\nu_2\nu_3 }^2}{\Gamma^2}\label{eq:gamdottot}\\
\frac{\dot{\gamma}(0,{\bf k}_f)}{N\Gamma^2}&=&-2+\sum_{\nu_1\nu_2\nu_3}W_{\nu_1\nu_2\nu_3}
\frac{\Gamma_{\nu_1\nu_2\nu_3 }}{\Gamma} \cos (\phi_{\nu_1\nu_2\nu_3 })
\label{eq:gamdotang}\\
W_{\nu_1\nu_2\nu_3}&=&\left( 1-\frac{|\nu_1 |}{N_1} \right)
\left( 1-\frac{|\nu_2 |}{N_1} \right)\left( 1-\frac{|\nu_3 |}{N_1} \right)
\end{eqnarray}
where both summations are for $-N_1<\nu_1,\nu_2,\nu_3<N_1$, the number of
atoms $N=N_1^3$,
$\phi_{\nu_1\nu_2\nu_3 }={\bf k}_f\cdot (\nu_1{\bf a}_1 +\nu_2{\bf a}_2 +\nu_3{\bf a}_3 )$
and $\Gamma_{\nu_1\nu_2\nu_3 }=2{\rm Re}[g(\nu_1{\bf a}_1 +\nu_2{\bf a}_2
 +\nu_3{\bf a}_3)]$. The function $W_{\nu_1\nu_2\nu_3}$ is the weighting
 from the number of terms with differences $\nu_1,\nu_2,\nu_3$.
The extension to one- and two-dimensional arrays is straightforward:
restrict $\nu_2=\nu_3=0$ in one-dimension and restrict $\nu_3=0$
in two-dimensions.

This formulation, Eqs.~\eqref{eq:gamdottot} and \eqref{eq:gamdotang},
shows why the one-dimensional case converges to a finite value in
the limit $N\to\infty$. The scaled initial slope of the photoemission rate
goes to an asymptotic limit as $N\to\infty$:
\begin{eqnarray}
\lim_{N\to\infty}\frac{\dot{\gamma}(0)}{N\Gamma^2}&=&-2+\sum_{\nu=-\infty}^\infty\frac{\Gamma_{\nu }^2}{\Gamma^2}\\
\lim_{N\to\infty}\frac{\dot{\gamma}(0,{\bf k}_f)}{N\Gamma^2}&=&-2+\sum_{\nu=-\infty}^\infty\frac{\Gamma_{\nu }}{\Gamma}\cos (\phi_{\nu })
\end{eqnarray}
The first summation is absolutely convergent because $\Gamma_{\nu }^2$
is proportional to $1/\nu^2$ for large $|\nu |$. The second summation is
conditionally convergent because $\Gamma_{\nu }\cos (\phi_{\nu })$
is proportional to $1/|\nu |$ times an oscillating function of $\nu$ for
large $|\nu |$.

\begin{figure}
\resizebox{86mm}{!}{\includegraphics{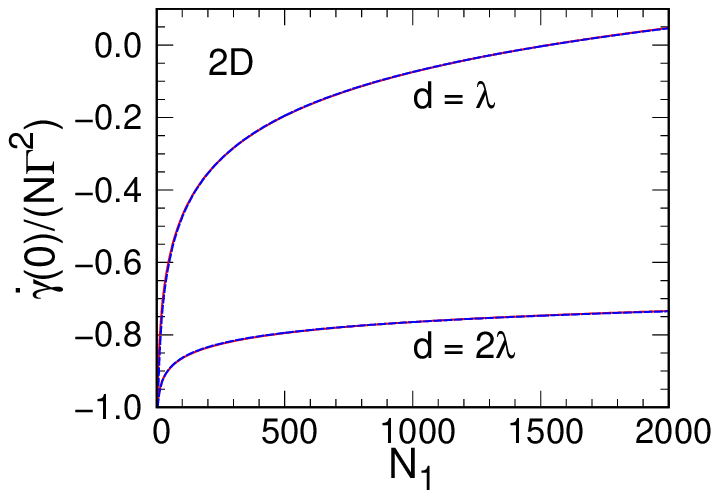}\includegraphics{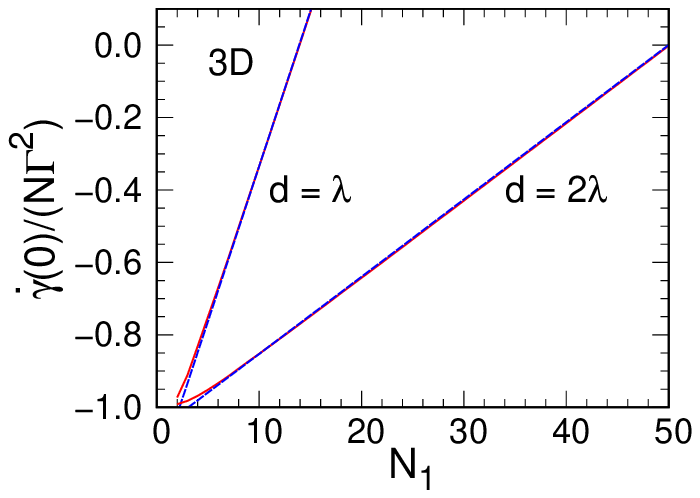}}
\caption{\label{fig:gamma_dot}
For the same case as Figs.~\ref{fig:2D_1plane} and \ref{fig:3D_1cube},
a plot of the scaled
initial slope of total radiation at a separation $d=\lambda$ and $2\lambda$
versus
the number of atoms in each direction. The red solid line is the calculation
using Eq.~\eqref{eq:gamdottot} and the blue dashed line is a plot
of $-1.2762 + 0.1740\ln{N_1}$ and $-1.0608+0.0429\ln{N_1}$ for two-dimensional
and $-1.1913+0.0856 N_1$ and $-1.0608+0.0213 N_1$ for three-dimensional.
The plane of atoms with polarization
perpendicular to the plane is superradiant for $N_1>1530$ for $d=\lambda$
while the cubic array is superradiant for $N_1\geq 14$.
}
\end{figure}

From the formulation, Eqs.~\eqref{eq:gamdottot} and \eqref{eq:gamdotang},
we can show the two-dimensional case always leads to superradiance in
the limit $N\to\infty$. For example, in Eq.~\eqref{eq:gamdottot},
the sum diverges proportional to $\ln{N_1}$ because the $\Gamma_{\nu_1\nu_2}
\propto 1/|{\bf a}_1\nu_1+{\bf a}_2\nu_2|$. As a specific example, a
square array with separation $d$ and polarization out of the plane gives
\begin{equation}\label{eq:2Dasym}
\frac{\dot{\gamma}(0)}{N\Gamma^2}\sim C+\frac{9\lambda^2}{32\pi^2d^2}\sum \frac{W_{\nu_1\nu_2}}{\nu_1^2+
\nu_2^2}\sim C + D\frac{\lambda^2}{d^2}\ln{N_1}
\end{equation}
in the limit of large $N_1$
with $C$ and $D$ constants. The asymptotic form was found by
converting the sum to an integral using $\nu^2=\nu_1^2+\nu_2^2$ 
with $\sum \to \int 2\pi \nu d\nu$. This can be seen in Fig.~\ref{fig:gamma_dot}
where we plot the scaled photon emission slope versus $N_1$. There is
a fit function which agrees well with the calculation with the fit being
$-1.2762 + 0.1740\ln{N_1}$ for $d=\lambda$ and
$-1.0608+0.0429\ln{N_1}$ for $d=2\lambda$.
This agrees better with Eq.~\eqref{eq:2Dasym} than might be expected given
the contribution from the weight function,
$9/(16 \pi ) = 0.1790$.

Unlike the one-dimensional case where the scaled slope
converges to a finite value as $N\to\infty$,
the two-dimensional scaled slope diverges for both
Eqs.~\eqref{eq:gamdottot} and \eqref{eq:gamdotang} implying there is always
a minimum number of atoms which will give superradiance. For
Fig.~\ref{fig:gamma_dot}, superradiance occurs for $N_1\simeq 1530$
corresponding to 2.3 million atoms for $d=\lambda$
and $N_1\sim 5.5\times 10^{10}$ corresponding to $N\sim 3\times 10^{21}$
atoms for $d=2\lambda$. These are very large numbers, not likely to be accessible
experimentally in the near future. The smallest $N_1$ for superradiance
is a rapidly increasing function of $d$: roughly
$N_1$ is squared for every increase of $d$ by a factor of $\sim\sqrt{2}$.
Also, the approximations in the basic
equations, Eq.~\eqref{eq:DenMat}, no longer hold when the array size
is too large so this asymptotic behavior only represents reality for
a finite range of $N_1$.

From the formulation, Eqs.~\eqref{eq:gamdottot} and \eqref{eq:gamdotang},
we can show that the three-dimensional case always leads to superradiance in
the limit $N\to\infty$. Following the logic of the two-dimensional
case, the
cubic array in $x,y,z$ with separation $d$ and polarization in the $z$-direction
gives
\begin{equation}\label{eq:3Dasym}
\frac{\dot{\gamma}(0)}{N\Gamma^2}\sim C + D\frac{\lambda^2}{d^2}N_1
\end{equation}
in the limit of large $N_1$ with $C$ and $D$ constants. The asymptotic form was found by
converting the sum to an integral using $\nu^2=\nu_1^2+\nu_2^2+\nu_3^2$ 
with $\sum \to \int 4\pi \nu^2 d\nu$.
This form can be seen in Fig.~\ref{fig:gamma_dot}
where we plot the scaled photon emission slope versus $N_1$. For $N_1$
greater than about 4, the scaled photoemission slope is proportional to $N_1$.
There is
a fit function which agrees well with the scaling with the fit being
$-1.1913+0.0856 N_1$ for $d=\lambda$ and
$-1.0608+0.0213 N_1$ for $d=2\lambda$. Note that the coefficient multiplying
the $N_1$ decreases by a factor of 4 in going from $d=\lambda$ to $2\lambda$
as expected from Eq.~\eqref{eq:3Dasym}.

Because the $\dot{\gamma}(0)/N$ increases more quickly with $N_1$ than
the two-dimensional case, the
region of superradiance is reached more quickly. For $d=\lambda$,
there is superradiance for $N_1\geq 14$ (corresponding to $N=2744$)
and, for $d=2\lambda$, it is
$N_1\geq 51$ (corresponding to $N=130,000$).
Compared to the two-dimensional case, these are much
smaller cutoff numbers although they are probably still experimentally
challenging in the near future.
The form of Eq.~\eqref{eq:3Dasym} suggests that the cutoff for superradiance
is $N_1=N^{1/3}\propto (d/\lambda)^2$.

\section{Summary}\label{sec:Summ}

We have presented an alternative method to Ref.~\cite{MAG2021} for determining
whether a collection of atoms will exhibit superradiance in the total emission
rate. Our condition is equivalent to that in Ref.~\cite{MAG2021} but uses the
trace of the square of a matrix instead of the variance of the eigenvalues.
We also presented a method for determining whether the collection of atoms
will exhibit superradiance only in particular directions. In addition to
expressions for fully inverted systems, we also found expressions for when the
gas is partially inverted in a product state. For the case of fully inverted
atoms, we determined the condition for superradiance for photoemission into more
than one final state. Finally, we showed how to efficiently evaluate these
expressions for arrays of atoms and determined the superradiance condition
for very large atom number.

For two level atoms,
we performed calculations for directional superradiance for one-, two-, and
three-dimensional arrays and found conditions of superradiance where the atom
separation was comparable to or larger than $\lambda$ for not very large numbers
of atoms. We showed that one-dimensional arrays have radiant properties that
converge to finite values as the number of atoms increase, but two- and
three-dimensional arrays have scaled radiant properties that increase with
increasing number of atoms. For fully inverted atoms, we showed how the decay
into many final states affects the superradiance condition for the total
emission rate.

While it will be difficult experimentally to have large, perfect arrays,
experiments with randomly situated atoms can be done. Effects that
result from the interference due to the perfect array will not be
reproduced in a random gas. However, the dependence of
the total photon emission rate with $d/\lambda$, $d$ the average separation,
and atom number, $N$, should be similar to that for a perfect
array when $N$ is large. For example, an effectively
two-dimensional cloud should have the scaled initial slope,
$\dot{\gamma}(0)/(N\Gamma^2)$, scale like Eq.~\eqref{eq:2Dasym}
and a three-dimensional cloud should scale like Eq.~\eqref{eq:3Dasym}.
They should {\it scale} like the perfect arrays because the main contribution
comes from atoms with separations $d\gg\lambda$ where
$\Gamma_{nm}^2$ does not vary strongly when averaged over
a wavelength.
This suggests that two- and three-dimensional gases also should
show superradiance and directional superradiance for enough atoms.

Data used in this publication is available at~\cite{data}.

{\it Note added}: After submission of this manuscript, we became aware
of related work in Ref.~\cite{SMA2021}

\begin{acknowledgments}
This work was supported by the National Science Foundation under Grant
No. 2109987-PHY.
\end{acknowledgments}

\appendix

\section{Equivalence of superradiance definition}\label{sec:AppEquiv}

This appendix shows that the condition $g^{(2)}(0)>1$ of Ref.~\cite{MAG2021}
is the same as
$\dot{\gamma}(0)>0$ from Eq.~\eqref{eq:Gdotfi}. From  Eq.~(12) in App.~B of
Ref.~\cite{MAG2021}, the superradiance condition is
\begin{equation}\label{eq:SupDef}
\sum_{\nu = 1}^N\Gamma_\nu^2-2N\Gamma^2>0
\end{equation}
with $\Gamma_\nu$ the eigenvalues of $\underline{\Gamma}$.
Since $\underline{\Gamma}$ is a real, symmetric
matrix, the sum of the squares of the eigenvalues
can be related to the trace of the squared matrix:
\begin{equation}
\sum_{\nu = 1}^N\Gamma_\nu^2 ={\rm Tr}[\underline{\Gamma}\; \underline{\Gamma}]
\end{equation}
Substituting this expression into Eq.~\eqref{eq:SupDef} immediately gives
\begin{equation}
{\rm Tr}[\underline{\Gamma}\; \underline{\Gamma}]-2N\Gamma^2=\dot{\gamma}(0)>0.
\end{equation}

\bibliography{super_rad_array}

\end{document}